\documentclass[a4paper]{jpconf}
\usepackage{graphicx}
\usepackage{multirow}
\usepackage{hhline}
\usepackage{hyperref}
\usepackage[caption=false]{subfig}
\usepackage[export]{adjustbox}[2011/08/13]

\begin{document}

\title{Optimizing ROOT IO For Analysis}

\author{B Bockelman$^1$, Z Zhang$^1$ and J Pivarski$^2$}

\address{$^1$ Holland Computer Center, University Nebraska $-$ Lincoln, Lincoln, NE 68588, USA}
\address{$^2$ Physics Department, Princeton University, Princeton, NJ 08544, USA}

\ead{bbockelman@cse.unl.edu}

\begin{abstract}
The ROOT I/O (RIO) subsystem is foundational to most HEP experiments - it provides a file format, a set of APIs/semantics, and a reference implementation in C++. It is often found at the base of an experiment's framework and is used to serialize the experiment's data; in the case of an LHC experiment, this may be hundreds of petabytes of files! Individual physicists will further use RIO to perform their end-stage analysis, reading from intermediate files they generate from experiment data.

RIO is thus incredibly flexible: it must serve as a file format for archival (optimized for space) and for working data (optimized for read speed). To date, most of the technical work has focused on improving the former use case. We present work designed to help improve RIO for analysis. We analyze the real-world impact of LZ4 to decrease decompression times (and the corresponding cost in disk space). We introduce new APIs that read RIO data in bulk, removing the per-event overhead of a C++ function call. We compare the performance with the existing RIO APIs for simple structure data and show how this can be complimentary with efforts to improve the parallelism of the RIO stack.

\end{abstract}

\section{Introduction}
\label{sec:introduction}

The field of High Energy Physics is unique in how uniformly the community has standardized on a foundational framework, ROOT. ROOT \cite{Brun199781} is an object-oriented C++ software framework originally developed at CERN.  While the framework integrates most tools a physicist may need - from mathematical functions, to an XML parser, to a GUI - one of the most frequently-used features is the C++ object serialization capability. This capability is used as the file format of hundreds of petabytes of HEP experiment data.  As nearly all experiments store their archival data in the format, it's also widely adopted by scientists producing their own derived data.  This produces a delicate balancing act: the ROOT IO format must simultaneously meet the needs of HEP experiments and their users:
\begin{itemize}
\item Experiments need features like schema evolution and the ability to serialize complex, arbitrary C++ objects.  Minimizing the storage footprint is a driving motivation (as they must manage tens of petabytes of data), and speed is low priority.
\item Analysts only need to serialize simple objects; speed is paramount and there is less sensitivity to overall file size.
\end{itemize}

As experimental storage costs are easier to quantify than time spent waiting for results, the priority has historically been skewed toward meeting the experimental needs.  In this paper, we explore new techniques to provide some specialization for analysis use cases in the ROOT framework.  We argue specialization for analysis is critical as users may iterate across their private datasets many times – and science can’t proceed until the IO has finished.

We introduce these specializations through revisiting some tradeoffs.  First, we explore a compression algorithm (LZ4) that has lower compression ratios but markedly faster decompression rates.  We feel this is justified as the input to a typical analysis can often fit on a single hard drive, so a modest increase in dataset size may be acceptable.  Second, we introduce the ``bulk IO" APIs, which provide access to many events per ROOT framework library call.  Here, the tradeoff is that the bulk IO drastically restricts the supported object types.  We view this as acceptable as analysis events are often drastically simplified when compared to full experiment frameworks.  Finally, we utilize some simple parallelism techniques to decompress file data in parallel even when the user may be only utilizing a single thread of control.


\section{Background}
\label{sec:background}

The vast majority of the data generated by a HEP experiment is \textit{event data}, which typically corresponds to a physical or simulated occurrence of particle collision.  Each event is decomposed into multiple C++ objects in ROOT that describe the understanding of the event. Within ROOT, an ordered list of event data is represented by the \texttt{TTree}.  The \texttt{TTree} is further partitioned into branches called \texttt{TBranch}; each branch collects a set of similar objects from events (typically, objects of the same C++ type).  For a more thorough discussion of the file organization, see \cite{Brun199781}.

When a C++ object is serialized, the resulting byte stream is stored in a memory buffer. Each branch contains one memory buffer.  When the buffer is full, ROOT compresses the data and a \texttt{TBasket} is created. \texttt{TBasket} is derived from \texttt{TKey} and contains additional information specific to the \texttt{TTree} navigation logic.  Hence, the serialized event data in a \texttt{TTree} consist of a sequence of compressed \texttt{TBasket}s and their metadata information.

As each event can be variable size, each \texttt{TBasket} for a branch may correspond to a different event range.  In the worst case scenario, the contents of an event may be scattered across hundreds of megabytes of file data: this potentially-poor locality is addressed by creating \textit{event clusters} -- ranges at which all buffers are flushed to disk, even if the corresponding memory buffers are not full.


\section{Bulk IO}
\label{sec:bulkio}

The typical mechanism for iterating through data in a \texttt{TTree} is a handwritten for-loop: for each event in the tree, fetch the data for that event from ROOT into some proxy object(s) and execute a user-written code.

When the user code is computationally expensive, the cost of the library call to ROOT is amortized into effectively nothing.  However, as Figure \ref{fig:decompressinglz4} shows, as the (uncompressed) bytes read per event decreases below one kilobyte, the ROOT overheads first become measurable, then eventually dominates for smaller event sizes.  Accordingly, we introduce a new interface for ROOT to copy all events in an on-disk \texttt{TBasket} directly to a user-provided memory buffer.  For the simplest of cases - primitives and C-style arrays of primitives - where the serialization can be done without a separate buffer or ``fixing up" pointer contents, the user can request the serialized data be delivered to the buffer or deserialized data.  By requesting the serialized data directly and deserializing directly in the event loop, the user can avoid an expensive scan from main memory.

Pragmatically, the user will not implement code for deserializing data themselves: rather, we have provided a header-only C++ facade around the data, allowing the user to work with a proxy object.  This allows the compiler to inline the deserialization code in the correct place.

\begin{figure}[!ht]
\centering
\includegraphics[height=3.5in, width=6in]{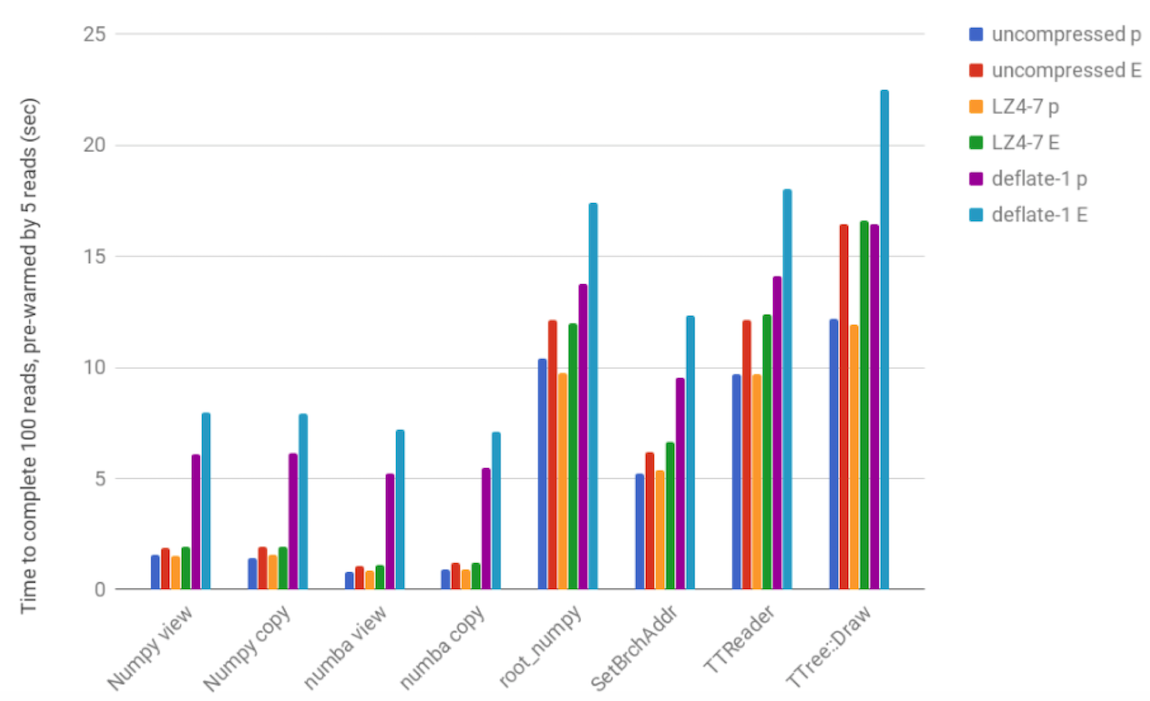}
\caption{Performance of Bulk IO.  Each access method was tested 6 times, using three different compression settings (uncompressed, LZ4, or ZLIB/deflate) and two different user calculations (momentum, $p$ and energy, $E$).  The energy calculation  utilizes baskets that are poorly aligned with event clusters; this measurement should illustrate the cost of the code for this special case}
\label{fig:bulkio}
\end{figure}

As is shown in Figure \ref{fig:bulkio}, this technique provides some drastic performance improvements.  The bulk IO APIs - even if accessed from within various Python-based mechanisms - are up to 10x faster in events per second compared to the traditional C++ mechanisms ROOT users are familiar with.

The file for this study is a flat ntuple of px, py, pz, and mass of dimuons. The baskets of the mass branch, however, are not aligned with the baskets of px, py, and pz, so total momentum can be calculated without copying any basket data while total energy must copy basket data to create aligned arrays. The distinction between ``viewing" and ``copying" is another layer of copying, this time defensively, since a view into unowned data is unsafe. The Numpy calculation produces intermediate arrays and performs one operation per pass over the data, while Numba computes the whole expression in one pass without unnecessary allocation. We see that this difference between Numpy and Numba is a 20\% effect, but that is dwarfed by differences with respect to {\tt SetBranchAddress}/{\tt GetEntry} and methods based on it, such as root\_numpy, {\tt TTreeReader}, and {\tt TTree::Draw}. Also, the differences are almost entirely washed out by deflate (ZLIB) decompression--- the value of BulkIO is only exposed by uncompressed and LZ4-compressed files.


\section{LZ4 Compression}
\label{sec:lz4}

By default, ROOT uses ZLIB \cite{Deutsch:1996:ZCD:RFC1950} as the default compression algorithm; it provides a balance between compression ratios and decompression speed. The alternate algorithm, LZMA \cite{lzma} is skewed toward a higher compression ratio and significantly slower decompression (detailed performance comparison is provided in \cite{DBLP:journals/corr/ZhangB17}). LZMA is ideal for the largest datasets (or those infrequently read and kept on tape) but, for data analysis, total dataset size may only be a few terabytes and reading speed is the critical metric. We explore another compression algorithm - LZ4 \cite{lz4} - which sacrifices compression ratio for improved read performance.

\begin{figure}[!ht]
\centering
\includegraphics[height=2.3in, width=3.1in]{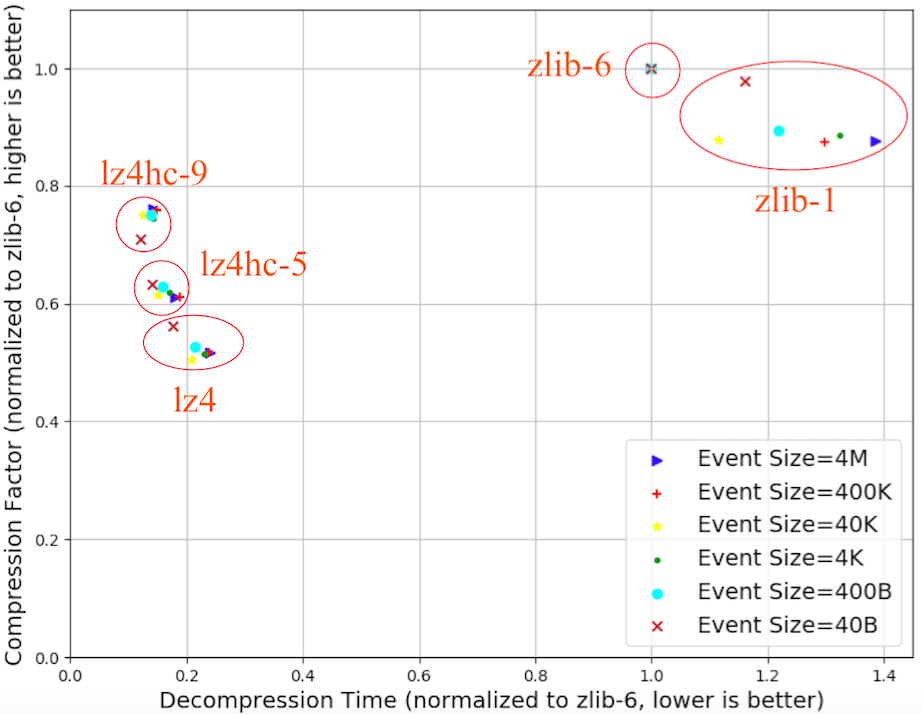}
\caption{Comparison of Different Compression Algorithms}
\label{fig:compressionalgorithms}
\end{figure}

Figure \ref{fig:compressionalgorithms} shows compression ratios and decompression speed for various of compression levels of ZLIB and LZ4. We normalize performance to ZLIB-6 (higher levels result in higher compression ratios). For LZ4's compression levels, \textbf{hc} represents a high-compression variant of the algorithm. In general, LZ4 has lwower compression ratios and faster decompression.

\begin{figure}[!ht]
\centering
\includegraphics[height=2in, width=4.4in]{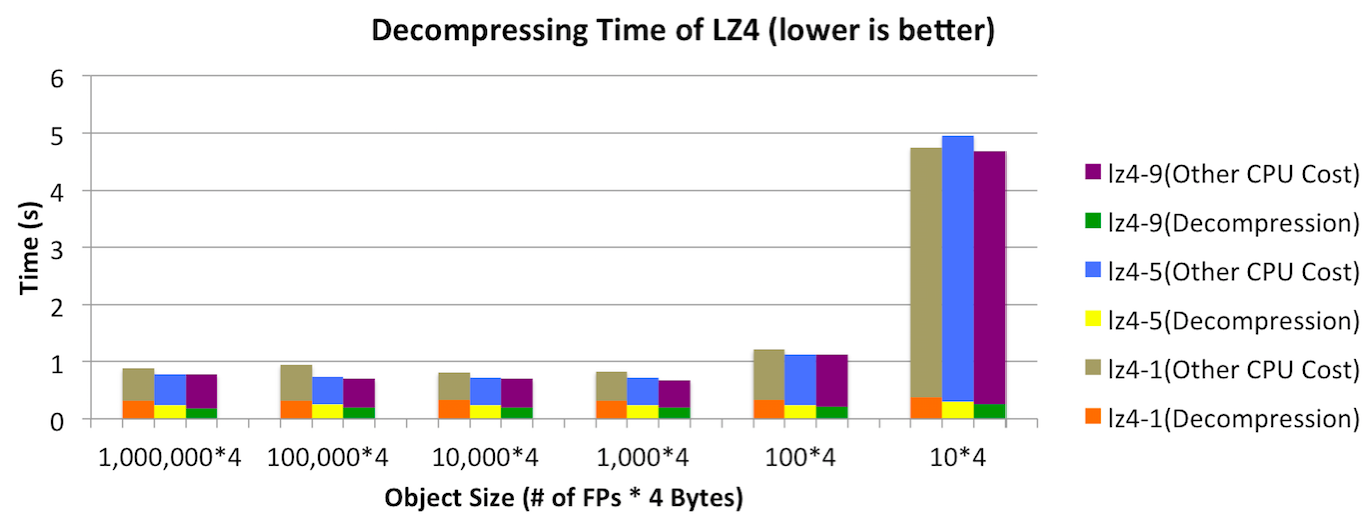}
\caption{CPU costs for reading LZ4-compressed files}
\label{fig:decompressinglz4}
\end{figure}

Decompression is not the only CPU cost when reading; for example, the uncompressed bytes must be converted to valid C++ objects. In Figure \ref{fig:decompressinglz4}, we measure the decompression time separately from other CPU costs. We create ROOT files with an aggregate of 400MB of event data, but varying individual event size from 40 bytes to 4MB. At one extreme, we create a ROOT file with 100 events, each containing 1,000,000 floating point (FPs) numbers. On the other END, we create 10,000,000 events and each event contains 10 FPs. As seen in the figure, while decompression time is roughly consistent across event size, other CPU costs come to dominate as the event size decreases.  Reading 10 small branches is not unreasonable, suggesting CPU overheads in ROOT IO for analysis (motivating the bulk IO work in Section \ref{sec:bulkio}).

\section{Asynchronous Parallel Unzipping}
\label{sec:parallelunzipping}

To accelerate reading for analysis, we leverage Intel Thread Building Blocks (TBB) \cite{tbb} to decompress multiple baskets in parallel. TBB allows users to break down high-level jobs into tasks that can run in parallel, allowing the library to manage and schedules threads for task execution.  When ROOT encounters a new event cluster, we create one task per approximately 100KB of compressed baskets, returning control to the calling thread as soon as possible.  The calling thread only blocks if the user requests event data whose decompression is not yet complete.

\begin{figure}[!ht]
\centering
\includegraphics[height=2.4in, width=3.6in]{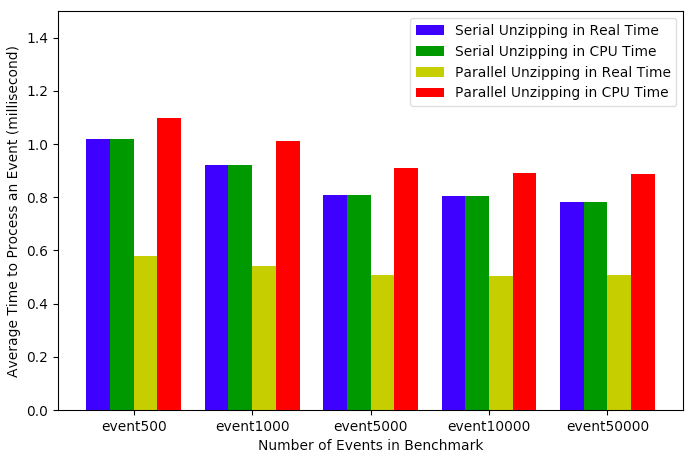}
\caption{Performance between Serial and Parallel Unzipping on Event Benchmark}
\label{fig:unzippingevent}
\end{figure}

In Figure \ref{fig:unzippingevent}, we test decompressing speed on synthetic event benchmark. We create ROOT files with 500 - 50,000 events. The performance is measured by sequentially reading all events. The figure shows the average runtime per events, running on a desktop-class machine with Intel 4-core CPU and Ubuntu 14.04.

As background threads can decompress data prior to the main thread accesses it, Figure \ref{fig:unzippingevent} parallel unzipping improves read performance, with more noticeable improvements at higher event counts. Parallel unzipping takes 52\% - 58\% processing time of serial unzipping. These synchronization techniques  require CPU cycles; this technique takes 8\% - 13\% more CPU cycles.


\section{Conclusions}

Analysis of ROOT data is a distinct use case from long-term archival - or data reconstruction - requiring fast IO with less sensitivity to compression ratio. In this paper, we discussed various ways to improve the read performance. For simple data types, bulk IO allows deserialization without multiple library calls.  The LZ4 algorithm provides significantly faster decompression than the existing ZLIB. Finally, the asynchronous parallel unzipping baskets can further double the read performance comparing to the serial baseline.

\section*{Acknowledgments}

This work was supported by the National Science Foundation under Grant ACI-1450323. This research was done using resources provided by the Holland Computing Center of the University of Nebraska.

\end{document}